\documentclass[fleqn]{another}
\input{epsf.sty}
\begin{document}

\begin{article}

\begin{opening}

\title{Estimations of orbital parameters of
exoplanets from transit photometry 
by using dynamical constraints}
\author{Zsolt S\'andor}
\runningauthor{Zs. S\'andor}
\runningtitle{Estimation of orbital parameters of exoplanets}
\institute{Department of Astronomy, E\"otv\"os University \\
H-1117 Budapest, P\'azm\'any P\'eter stny. 1/A\\
\texttt{e-mail: Zs.Sandor@astro.elte.hu}}

\begin{abstract}
The probability of the detection of Earth-like exoplanets may increase in the near future after the launch of the space missions using the transit photometry as observation method. By using this technique only the semi-major axis of the detected planet can be determined, and there will be no information on the upper limit of its orbital eccentricity. However, the orbital eccentricity is a very important parameter, not only from a dynamical point of view, since it gives also information on the climate and the habitability of the Earth-like planets. In this paper a possible procedure is suggested for confining the eccentricity of an exoplanet discovered by transit photometry if an already known giant planet orbits also in the system. 
\end{abstract}


\end{opening}

\section{Introduction}
In the last decade, following the discovery of the first extrasolar planet around 51 Pegasi \cite{Mayor-Queloz:NAT95}, more than 168 exoplanets have been observed. The detection of exoplanets is of great importance, since they form planetary systems around their hosting stars, and by studying the main properties of these exoplanetary systems, the characteristics, formation and evolution of the Solar System can be treated and understood from a wider perspective. However, the above picture is rather ideal than complete yet, since the exoplanets observed until now are mainly Jupiter-like gas giants. This is a consequence of the fact that by using radial-velocity measurements, which is the most effective ground-based observing technique, it is not possible to detect Earth-like planets yet. (We note, however, that the detection of a $\sim 8 M_{\oplus}$ planet by Rivera et al. (2005) indicates that radial velocity searches might be able to detect Earth-like planets in the future.)

The most compelling question related to exoplanetary research is the detection of Earth-like planets. Beside their importance in testing and improving formation theories of planetary systems, a major question is their habitability. If an Earth-like planet orbits in the habitable zone of a star, there could be chances of appearing (water based) life on its surface. On the habitable zone we mean that region around a star, where liquid water can exist on the surface of a planet, for details see in \cite{Kasting-etal:ICAR93}.

In order to find Earth-like planets, there are space missions in construction and planning phase. Such a mission is COROT (sponsored by CNES, ESA and other countries) to be launched in 2006, and the Kepler Mission (NASA) with a launch in 2008. These missions will use the transit photometry as detection technique, which is based on measuring the periodic dimming of a star's light intensity caused by an unseen transiting planet. Observations performed by these instruments will provide the semi-major axis $a$ of the transiting planet calculated from Kepler's third law by measuring the period $T$ of the transits, and assuming that the mass $m_*$ of the hosting star is known.  Since $m_*$ is known only with limited accuracy, an uncertainty appears in $a$. (We note that $m_*$ can be determined by spectroscopic observations and by stellar model calculations.) The error appearing in $a$ and in the eccentricity $e$ of the transiting planet due to the uncertainties in stellar mass will be estimated in 3.2.

In this paper we present a procedure which can be used to estimate the orbital eccentricity and inclination of a transiting planet, if (i) we can measure the duration of the transit, and (ii) there is another (giant) planet in the system. We shall see that an equation can be derived, which connects the mass and the radius $R$ of the star, the semi-major axis $a$, the eccentricity $e$, the argument of the periastron $\omega$, the inclination $i$ of the transiting planet, and the duration $\tau$ of the transit. In this equation there are three unknowns, $e$, $\omega$, and $i$. By fixing $i$, the corresponding $(\omega,e)$ pairs can be visualized as curves on the $\omega-e$ parameter plane. Thus the problem is underdetermined and there is no way to give an upper limit for $e$. We note that during one revolution of the transiting planet, in principle there are two minima in the {\it light curve} of the star. The first minimum occurs when the planet moves in front of the star, the second appears when the star is between the observer and the planet. The maximum of the light curve can be observed when both the star and the planet are visible. By using the second minimum observation, the eccentricity of the transiting planet can be estimated. However, the maximum and the second minimum of the light curve can not be seen in the case of an Earth-like planet, since its contribution to the whole light flux is not detectable. (The light curve is the variation of the light intensity of the star as a function of the orbital phase of the transiting planet.)

On the other hand, as suggested by planetary formation scenarios, we expect that beside Earth-like planets Jupiter-like giant planets can also be found in the majority of planetary systems. Having discovered an Earth-like planet around a star, by using complementary techniques (as observations by Space Interferometry Mission and ground-based Doppler spectroscopy) other more massive planets can be identified in the system, and their orbital parameters can be determined too. (Although the above scenario sounds very optimistic, it is one of the scientific goals of Kepler Mission, see\\ \texttt{http://kepler.nasa.gov/basis/goals.html}.) 

The presence of one or more additional giant planets (beside the transiting one) results in that, that both ordered and chaotic regions appear in the phase space of the system. If the phase trajectory of the Earth-like planet is in an ordered region of the phase space, the motion of the planet is stable for arbitrary long times. If the initial conditions of its orbit are in a chaotic region of the phase space, the motion of the planet can become unstable after some time. In this paper we try to exclude those orbital parameters of the transiting planet, which result in chaotic motion. We shall demonstrate that in some cases it is possible to determine an upper limit for the eccentricity and a lower limit for the inclination of the transiting planet. We stress that the eccentricity is a very important orbital parameter not only from a dynamical point of view, but also in studying the habitability and climatic variations of the Earth-like planet.  

The paper is organized as follows: first we derive an equation between the duration of the transit and some important parameters of the star and the transiting body, then we solve this equation numerically. After examining the solutions of this equation, we map the stability structure of the system assuming the presence of a giant planet. Then we determine lower limits for the inclination and an upper bound for the eccentricity of the transiting planet depending on the eccentricity and the semi-major axis of the known giant planet.

\section{An equation connecting the parameters of a star, a transiting planet and the transit}
In this section we shall derive an equation between the orbital parameters of the transiting planet, the star's mass, and the duration of the transit from the geometry of the transit. 

\begin{figure}[t]
\centering{\epsfysize=5cm \epsfbox{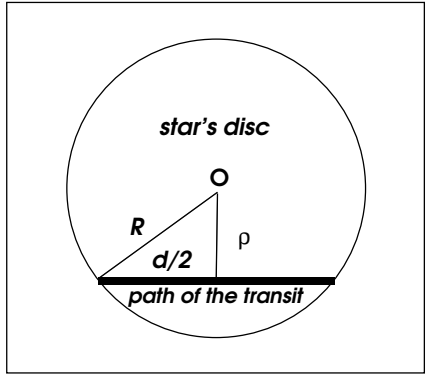}}
\caption{The transit of a planet in front of the stellar disc. The straight sections denoted by $R$, $\rho$, and $d/2$ form a pythagorean triangle.}
\label{fig1}
\end{figure}

\begin{figure}[h]
\centering{\epsfysize=5cm \epsfbox{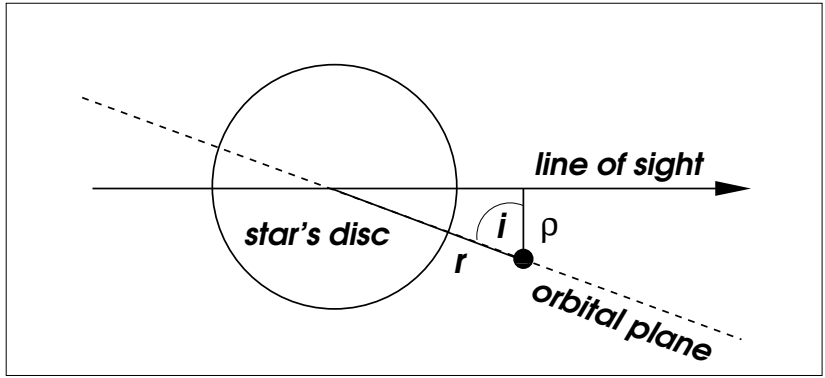}}
\caption{Side-view of the transit, where $r$ is the distance of the planet from the star's center and $i$ is the inclination of its orbital plane.}
\label{fig2}
\end{figure}

Let us suppose that the star's disc is a circle with radius $R$, and a planet is moving in a front of this disc with an average velocity $v_{\mathrm{tr}}$. If the duration of the transit is denoted by $\tau$ and the lenght of the path of the transiting planet is $d$ (see Figure \ref{fig1}), the following approximation holds:
\begin{equation}
 v_{\mathrm{tr}} = \frac{d}{\tau} .
 \label{eq1}
\end{equation}
We note that according to Kepler's second law, the velocity of the planet is changing during the transit (except in the case of circular orbits), however this change is negligible for moderate eccentricities. 
From the triangle in Figure \ref{fig1}
\begin{equation}
R^2=\rho^2+\left( \frac{d}{2}\right) ^2 .
\label{eq2}
\end{equation}
From Equation (\ref{eq2}) the lenght of the transit's path $d$ can be expressed as
\begin{equation}
d=2\sqrt{R^2-r^2\cos^2 i},
\label{eq3}
\end{equation}
where, according to Figure \ref{fig2}, 
\begin{equation}
\rho=r\cos i,
\label{eq4}
\end{equation}
$i$ being the inclination (e.g. the angle between the orbital plane and the tangent plane to the celestial sphere), and $r$ the distance between the center of the star and the planet at the moment of the middle of the transit. 

\begin{figure}[h]
\centering{\epsfysize=5cm \epsfbox{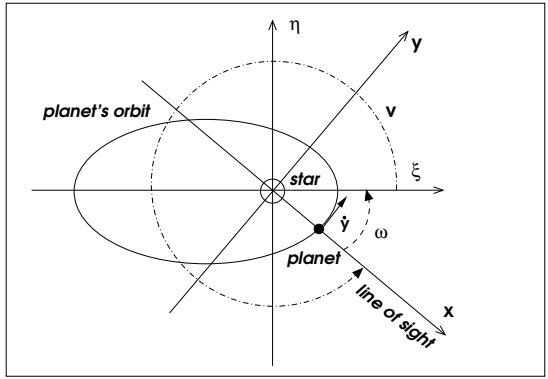}}
\caption{The transit as viewed from above. At the middle of the transit $v_{tr}$ is nearly equal to $\dot y$. The coordinate system $(\xi,\eta)$ is the rotation of the coordinate system $(x,y)$ by $\omega$.}
\label{fig3}
\end{figure}

By using the well known formula for $r$:
\begin{equation}
r=\frac{a(1-e^2)}{1+e\cos v}\ ,
\label{eq5}
\end{equation}
(where $a$ is the semi-major axis, $e$ is the eccentricity, and $v$ is the true anomaly of the transiting planet), and Equations (\ref{eq1}) and (\ref{eq3}),  the average transiting velocity of the planet ($v_{tr}$) can be written as 
\begin{equation}
v_{\mathrm{tr}}=\frac{2}{\tau}\sqrt{R^2-\left[ \frac{a(1-e^2)}{1+e\cos v}\right] ^2\cos^2 i}\ ,
\label{eq6}
\end{equation}
where $v$ is the true anomaly of the transiting planet at the middle of the transit.

On the other hand, $v_{\mathrm{tr}}$ can also be approximated on the basis of the two-body problem. In the coordinate system $(\xi,\eta)$, in which the axes of the orbital ellipse are on the axes $\xi$ and $\eta$, the components of the orbital velocity vector are \cite{mur99}:
 \begin{eqnarray}
 \lefteqn{\dot\xi=-\sqrt{\frac{\mu}{p}}\sin v\ ,} \\ [10pt] \nonumber 
 \lefteqn{\dot\eta=\sqrt{\frac{\mu}{p}}(e+\cos v)\ ,}
 \end{eqnarray}
where $p=a(1-e^2)$ is the parameter of the ellipse and $\mu=k^2(m_*+m_p)$, $m_*$ and $m_p$ being the stellar and planetary masses respectively, and $k$ is the Gaussian constant of gravity. Let $(x,y)$ denote the coordinate system in which the $x$-axis is the projection of the line of sight (e.g. the line connecting the center of the star to the observer) to the orbital plane of the transiting planet. From Figure \ref{fig3} it can be seen that the system $(\xi,\eta)$ is just a rotation of the system $(x,y)$ by an angle $\omega$, which is the argument of the periastron of the transiting planet. Thus in the coordinate system $(x,y)$ formulae (7) transform as
 \begin{eqnarray}
 \lefteqn{\dot x = \dot\xi\cos\omega-\dot\eta\sin\omega\ ,}  \\ [10pt] \nonumber 
 \lefteqn{\dot y = \dot\xi\sin\omega+\dot\eta\cos\omega\ .}
 \end{eqnarray}
 
Studying Figure \ref{fig3} one can find that the average velocity $v_{tr}$ of the transiting planet is almost equal to $\dot y$, which is the velociy of the planet at the middle of the transit. Then by using Equations (7) and (8) we find
\begin{equation}
v_{\mathrm{tr}} \approx \dot y = -\sqrt{\frac{\mu}{p}}\sin v\sin\omega + \sqrt{\frac{\mu}{p}}(e+\cos v)\cos\omega.
\label{eq9}
\end{equation}
Studying again Figure \ref{fig3}, it is also true that at the middle of the transit 
\begin{equation}
v + \omega = 360^{\circ}\ ,
\label{eq10}
\end{equation}
thus the average orbital velocity of the transiting planet is
\begin{equation}
v_{\mathrm{tr}} = \sqrt{\frac{\mu}{p}}(1 + e\cos\omega)\ .
\label{eq11}
\end{equation}

Combining Equations (\ref{eq6}), (\ref{eq10}), and (\ref{eq11}) we obtain a relation between the orbital parameters of the transiting planet and the duration of the transit:
\begin{equation}
\sqrt{\frac{\mu}{a(1-e^2)}}(1 + e\cos\omega) - \frac{2}{\tau}\sqrt{R^2-\left[ \frac{a(1-e^2)}{1+e\cos \omega}\right] ^2\cos^2 i} = 0.
\label{connectingeq}
\end{equation}
This equation has the form
\begin{equation}
f(a,e,i,\omega,\mu,R,\tau)=0\ ,
\end{equation}
where the unknown quantities are the eccentricity $e$, the inclination $i$, and the argument of the periastron $\omega$. The other quantities, such as the semi-major axis $a$, the mass parameter ($\mu$), the radius of the star ($R$), and the duration of the transit ($\tau$) are supposed to be known.

\section{Analysis of Equation (12)}
In this section we present and analyse the solutions of Equation (\ref{connectingeq}) by using a model system. We shall see how the orbital eccentricity $e$ of the transiting planet depends on its argument of periastron $\omega$, or in other words, on the orbital position of the planet during the transit. We also study the sensitivity of the solutions to the uncertainties of the mass of the hosting star.

\subsection{Solution of Equation (12)}
According to the last paragraph of the previous section, the unknown quantities in Equation (\ref{connectingeq}) are the inclination $i$, the argument of periastron $\omega$, and the eccentricity $e$ of the transiting planet. Thus by fixing $i$, Equation (\ref{connectingeq}) can be solved numerically, and the ($\omega,e$) pairs of the solution can be represented as a curve on the $\omega-e$ parameter plane. 

In order to study the solutions of Equation (\ref{connectingeq}), we give specific values for the parameters in Equation (\ref{connectingeq}). Let us assume that the mass of the transiting planet is 1 Earth-mass, and it revolves around a 1 Solar-mass star with radius $R = 6.96\times 10^8$ m, in an elliptic orbit characterized by $a=1$ AU, $e=0.1$, $i=89.95^\circ$. We consider two cases: (i) $\omega=30^\circ$ and (ii) $\omega=130^\circ$. It can be calculated easily that in these cases the durations of the transits are $\tau=0.488029$ day and $\tau=0.563743$ day, respectively.

\begin{figure}[h]
\centering{\epsfysize=5cm \epsfbox{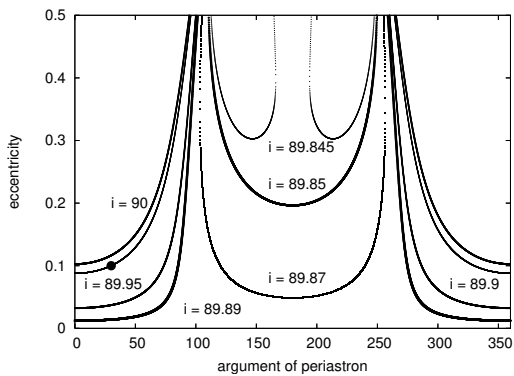}}
\caption{Solutions of Equation (\ref{connectingeq}) for different inclinations when $\tau=0.488029$ day. The original solution, which results in the above $\tau$, is marked with a filled circle at $\omega=30^\circ$, $e=0.1$, and $i=89.95^\circ$.}
\label{fig4}
\end{figure}
\begin{figure}[h]
\centering{\epsfysize=5cm \epsfbox{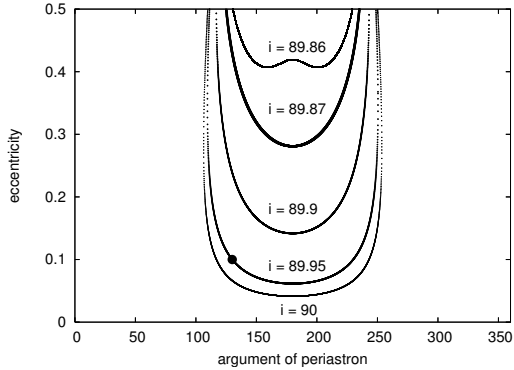}}
\caption{Solutions of Equation (\ref{connectingeq}) for different inclinations when $\tau=0.563743$ day. The original solution, which results in the above $\tau$, is marked with a filled circle at $\omega=130^\circ$, $e=0.1$, and $i=89.95^\circ$.}
\label{fig5}
\end{figure}

By observing transits caused by the above planet, we can measure their duration $\tau$ and period $T$, from which the semi-major axis $a$ can be calculated. In the first case $\tau=0.488029$ day, and for different values of $i$ the corresponding $\omega-e$ curves are plotted in Figure \ref{fig4} for $e<0.5$. (We note that it is possible to study larger values of $e$ as well, but for our demonstrational purpose we restrict ourselves for $e<0.5$.) We also mark the real $(\omega,e)$ solution by a filled circle on the curve corresponding to $i=89.95^\circ$, but as we can see, there is no way to restrict efficiently the infinite set of solutions. The only restriction is that the solutions can not be chosen from the region above the $\omega-e$ curve corresponding to $i=90^\circ$. 

In the second case corresponding to $\tau=0.563743$ day, the $\omega-e$ curves are plotted in Figure \ref{fig5}. It can be seen that if the transit happens nearly at the apastron, the set of the solutions of Equation (\ref{connectingeq}) is more limited than in the first case. Only those $(\omega-e)$ pairs satisfy Equation (\ref{connectingeq}), which are in the region above the $i=90^\circ$ curve. The real solution is also marked by a filled circle on the curve $i=89.95^\circ$.  

Equation (\ref{connectingeq}) has an infinite set of solutions formed by pairs of $(\omega, e)$ values. If only the duration of the transit is known, it is not possible to choose which $(\omega,e)$ pair represents the real parameters of the transiting planet.  

\subsection{Sensitivity of the solutions of Equation (12) to the stellar mass' error}
As mentioned in the Introduction, the mass of the hosting star is known only with limited accuracy. This uncertainty in the stellar mass affects the semi-major axis of the transiting planet and causes an error in the eccentricity estimates. The semi-major axis of the transiting planet can be calculated from Kepler's third law 
\begin{equation}
\frac{a^3}{T^2}=\frac{k^2}{4\pi^2}(m_*+m_p), 
\label{kepler3}
\end{equation}
where $T$ is the period of the transits, $m_*$ is the mass of the hosting star, and $m_p$ is the mass of the transiting planet, respectively ($k$ is the Gaussian gravitational constant). In the case of Earth-like planets $m_p<<m_*$, so $m_p$ can be neglected. Denoting the errors in $a$ and $m_*$ by $\delta a$ and $\delta m_*$, respectively, and expanding Equation (\ref{kepler3}) around $a$ and $m_*$ up to first order we obtain:
\begin{equation}
3a^2\delta a = \frac{k^2 T^2}{4\pi^2} \delta m_* .
\label{expansion}
\end{equation}
Dividing Equation (\ref{expansion}) by $a^3$ and using again Equation (\ref{kepler3}) (by neglecting $m_p$) we find for the relative errors
\begin{equation}
3 \frac{\delta a}{a} = \frac{\delta m_*}{m_*}.
\label{errorseq}
\end{equation}
By using Equations (\ref{errorseq}) and (\ref{connectingeq}) we calculated numerically the relative errors in the eccentricity of the transiting planet as a function of $\omega$ for the specific case when $i=89.95^{\circ}$ and $\tau = 0.488029$ day. Then we determined the maximum values 
$(\delta e/e)_{\mathrm{max}}$ of the relative error in eccentricity for different values of $\delta m_*/m_*$. We found that for $\delta m_*/m_* = 3\%$  $(\delta e/e)_{\mathrm{max}} = 12\%$, for $\delta m_*/m_* = 6\%$ $(\delta e/e)_{\mathrm{max}}=25\%$, and for $\delta m_*/m_* = 10\%$ $(\delta e/e)_{\mathrm{max}}=38\%$. 

\section{A possible confinement of the eccentricity of the transiting planet}
\subsection{The case of one additional giant planet}
In this section we shall investigate the case when, beside the newly discovered planet, an already known giant planet also orbits around the hosting star. The presence of such a planet makes the problem non-integrable and both ordered and chaotic regions can be found in the phase space of the system. We suppose that the most probable orbital solutions of the transiting planet are those, which emanate from the ordered regions of the phase space. Such values of the orbital parameters of the transiting planet which would result in chaotic behaviour are unlikely, since in long terms the orbit of the planet could be unstable, thus these solutions should be avoided. We expect that the presence of a second (giant) planet represents a dynamical constraint, which reduces the infinit set of solutions of Equation (\ref{connectingeq}), and gives an upper limit for the maximum eccentricity of the transiting planet. We shall also demonstrate that by studying the solution-curves of Equation (\ref{connectingeq}) together with the stability structure of the $\omega-e$ plane, a lower bound for the inclination can also be determined. In what follows we shall investigate the stability in the $(\omega-e)$ plane within the framework of the planar general three-body problem. 

In order to map the stability properties of the $(\omega-e)$ plane we used the Relative Lyapunov Indicator (RLI) (S\'andor et al. 2000, 2004). The initial $\omega$ and $e$ values are chosen from the intervals $e\in [0,0.5]$ and $\omega\in [0^{\circ},360^{\circ}]$ with $\Delta e=0.025$ and $\Delta\omega=2^{\circ}$. The initial value of the semi-major axis of the transiting planet is always $a=1$ AU, while its true anomaly is calculated according to Equation (\ref{fig10}) as $v=360^{\circ}-\omega$ (see also Figure \ref{fig3}). The masses of the planets are 1 Jupiter-mass and 1 Earth-mass, respectively, and the mass of the hosting star is 1 Solar-mass.

For each pair of the initial $(\omega,e)$ values we assign the RLI of the corresponding orbit calculated for 500 periods of the transiting planet. If the RLI is small ($\sim 10^{-12}-10^{-13}$), the corresponding orbit is ordered and stable for very long time. If the RLI $\sim 10^{-11}-10^{-9}$ the orbit is weakly chaotic. In practical sense such an orbit could be (Nekhoroshev) stable for very long terms as well, however, it can not be stable for arbitrary long time. Thus the regions characterized by these RLI values can already be the birth places of unstable orbits. Orbits having larger RLI $\sim 10^{-8}-10^{-5}$, are strongly chaotic, and they will be unstable after certain time. In our stability maps the ordered regions are denoted by light, the weakly chaotic regions by grey, and the strongly chaotic regions by dark shades.   

\begin{figure}[!t]
\centering{\epsfysize=7cm \epsfbox{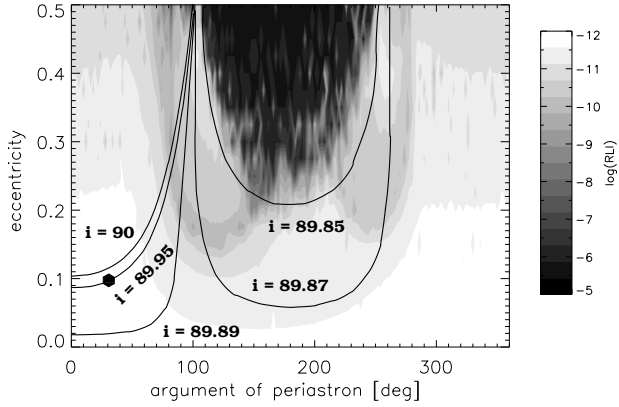}}
\caption{The stability map of the $\omega-e$ parameter plane, when $a_1=2.0$ AU and $e_1=0.2$. The $\omega-e$ curves for different $i$ are also plotted when $\tau=0.488029$ day.}
\label{fig6}
\end{figure}

\begin{figure}[!]
\centering{\epsfysize=7cm \epsfbox{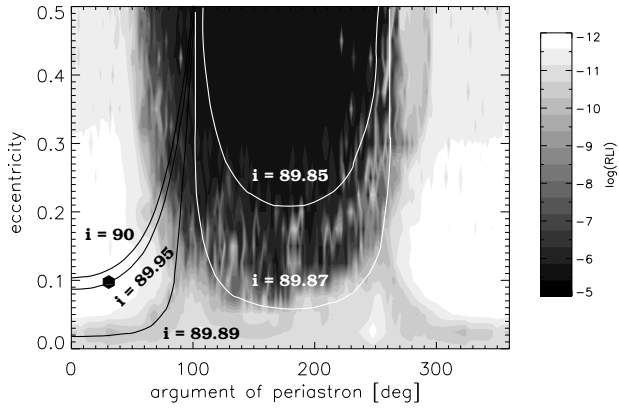}}
\caption{The stability map of the $\omega-e$ parameter plane, when $a_1=2.0$ AU and $e_1=0.3$. The $\omega-e$ curves for different $i$ are also plotted when $\tau=0.488029$ day.}
\label{fig7}
\end{figure}

In what follows we consider the cases where the parameters of the known giant planet having 1 Jupiter mass are the following: $a_1=2.0$ AU, $e_1=0.2$, and $0.3$. We fix the angular elements of the giant planet to $\lambda=\omega=0^\circ$. In Figures \ref{fig6} and \ref{fig7} we show the dynamical structure of the $\omega-e$ parameter plane for the two different values of $e_1$. In these figures we also plot the solution curves of Equation (12) by using $\tau=0.488029$ day.
From Figure \ref{fig6} it can be seen that there are two upper bounds for the eccentricity of the transiting planet depending on whether the transit occurs near the periastron, or near the apastron. If the transit is near the periastron $\omega<80^\circ$, the upper limit of the eccentricity is $e<0.27$, since the $\omega-e$ curves cross the chaotic region around this value. If the transit happens around the apastron $\omega\in[150^\circ,220^\circ]$, the upper limit of the transiting planet's eccentricity is $e<0.22$. 
In this case a lower limit can also be given for the inclination, $i>89.^\circ 85$.
The real solution is marked (as a filled circle) on the curve corresponding to $i=89.95^\circ$.

If the eccentricity of the giant planet is $e_1=0.3$, see Figure \ref{fig7}, the maximum upper limit of the transiting planet's eccentricity is $e<0.18$. However, in this case there exists a lower limit $e>0.05$ as well. If the transit took place around the periastron the corresponding $\omega$ and $e$ values would result in weakly chaotic orbits. A lower bound of the inclination in this case is $i>89.89^\circ$. Among the two values of the giant planet's eccentricity, this latter represents a more effective dynamical constraint for the orbital parameters of the transiting planet, which are $a=1.0$ AU, $e=0.1$, $\omega=30^\circ$, and $i=89.95^\circ$.

Next we shall study the cases when the transit takes place around the apastron $\omega=130^0$, resulting in a transit's duration $\tau=0.563743$ day. We used the same orbital parameters of the giant planet as in the previous case, $a_1=2.0$ AU, $e_1=0.2$ and $0.3$ respectively. 
\begin{figure}[!t]
\centering{\epsfysize=7cm \epsfbox{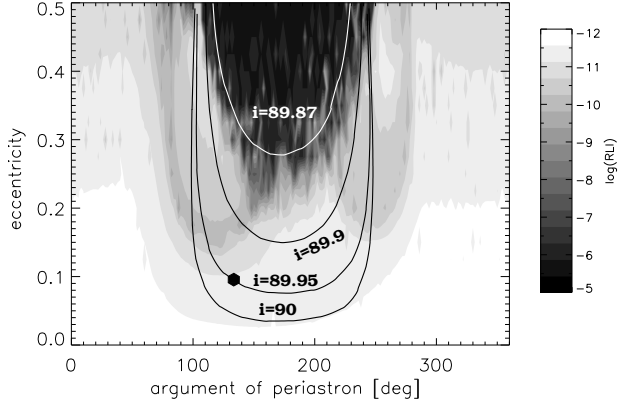}}
\caption{The stability map of the $\omega-e$ parameter plane, when $a_1=2.0$ AU and $e_1=0.2$. The $\omega-e$ curves for different $i$ are also plotted when $\tau=0.563743$ day.}
\label{fig8}
\end{figure}
\begin{figure}[!h]
\centering{\epsfysize=7cm \epsfbox{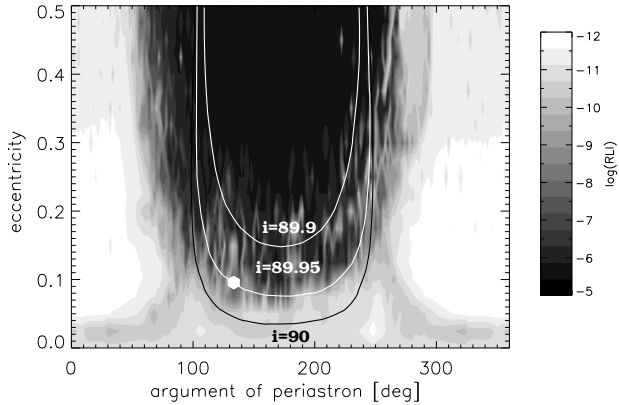}}
\caption{The stability map of the $\omega-e$ parameter plane, when $a_1=2.0$ AU and $e_1=0.3$. The $\omega-e$ curves for different $i$ are also plotted when $\tau=0.563743$ day.}
\label{fig9}
\end{figure}
Figure \ref{fig8} shows the situation when $e_1=0.2$. The maximum upper limit of the eccentricity of the transiting planet is $e<0.24$ corresponding to $\omega \approx 220^\circ$. The lowest bound of its inclination is $i>89.89^\circ$. 

Finally, when $e_1=0.3$, as shown in Figure \ref{fig9}, there are no stable solutions for the orbital parameters of the transiting planet. 

We have also investigated cases, when the semi-major axis of the giant planet was smaller or larger than $2$ AU. If $a_1$ is smaller, a smaller $e_1$ is enough to result in an effective dynamical constraint. If $a_1$ is larger, the eccentricity of the giant planet should be larger as well for an efficient dynamical constraint. 

In the above cases we assumed that the giant planet and the transiting planet revolved in the same plane. This, based on Solar System examples, seems to be a reasonable assumption, however, the orbital planes of the planets may differ slightly from each other. Thus we performed numerical simulations by using a small mutual inclination $I=5^{\circ}$ between the orbital planes as well. Comparing the corresponding $\omega-e$ parameter planes we have not found any significant differences between the planar and the spatial cases with small mutual inclination.

\subsection{The case of two additional giant planets}
Among the extrasolar giant planets with long observational baselines, there is a high rate of multiple planet systems. We expect that this rate will increase further with the accumulation of the observational data. Therefore we studied a case, when beside a transiting Earth-like planet there are two additional giant planets. 

\begin{figure}[!t]
\centering{\epsfysize=7cm \epsfbox{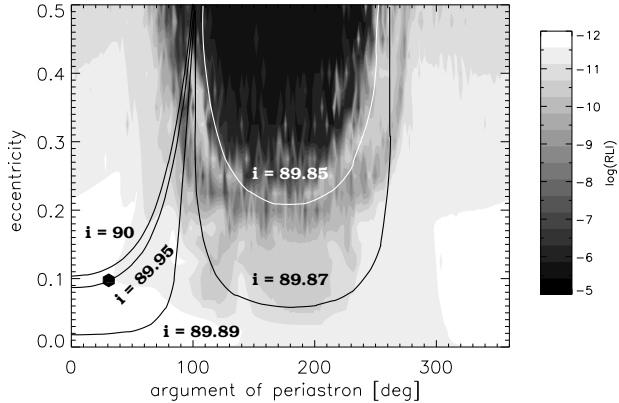}}
\caption{The stability map of the $\omega-e$ parameter plane, when two giant planets are in the system. The semi-major axes and the eccentricities of the giants are $a_1=2.0$ AU, $e_1=0.2$, $a_2=4.0$ AU, and $e_2=0.05$. The $\omega-e$ curves for different $i$ are also plotted when $\tau=0.488029$ day.}
\label{fig10}
\end{figure}

The inner giant planet's initial orbital parameters are the same as in 4.1. The outer giant planet, having 1 M$_{\mathrm{Jup}}$, moves in an orbit characterized by $a_2=4$AU, $e_2=0.05$ in the same plane as the transiting and the inner giant planet. Figure \ref{fig10} shows the stability properties of the $\omega - e$ parameter plane of the transiting planet. In order to study the effect of the outer giant planet on the $\omega - e$ parameter plane, we should compare Figure \ref{fig10} either to Figure \ref{fig8} or to Figure \ref{fig6}. In these figures the initial conditions for the Earth-like and the inner giant planet are the same. 

The main structure of these figures is very similar to each other, however in the case of the additional outer giant planet the strongly chaotic region on the $\omega - e$ parameter plane is more enhanced. This means that the dynamical constraint on the transiting planet could be more efficient in the case of two giant planets. Of course the strongness of the dynamical constraint depends on many parameters of the outer giant planet. To explore the complete parameter space is beyond the scope of this paper, it will be the subject of a forthcoming research.  
 
\section{Conclusions}
The detection of Earth-like extrasolar planets by using ground based spectroscopic methods is beyond the present capabilities of observational astronomy. In the near future space instruments will be launched such as COROT and Kepler Mission which are devoted to observe such planets by using transit photometry. 

In this paper we addressed the question whether it is possible to determine the orbital elements of Earth-like planets discovered by transit photometry if, apart from the period, the duration of the transit can be measured too. We supposed that the mass and the radius of the hosting star is known. We derived an equation, which connects the stellar and planetary masses, the duration of the transit, the semi-major axis, the eccentricity, the argument of periastron and the inclination of the transiting planet. By fixing the inclination, this equation contains two unknown variables, the argument of periastron $\omega$ and eccentricity $e$ of the transiting planet. Thus the solutions for different inclinations can be represented as curves on the $\omega-e$ parameter plane. 

In the last section of the paper we assumed that beside the transiting Earth-like planet additional giant planets orbit around the star as well. This assumption is quite reasonable if we accept the formation theories of planetary systems supporting the simultaneous presence of both rocky, Earth-like and gaseous, Jupiter-like planets. Since the detection of giant planets is possible by radial velocity measurements, we assumed their orbital parameters to be known. (We note that from radial velocity measurements only $m_p\sin i_p$ is known, where $m_p$ and $i_p$ are the mass and inclination of the giant planet. However, based on Solar System examples, we can assume that the giant planets orbit nearly in the same plane as the transiting Earth-like planet, so its inclination is $i_p\approx 90^\circ$.) By using the framework of the general three-body problem, we investigated the influence of the known giant planet(s) on the transiting Earth-like planet's $\omega-e$ parameter plane. We found that on the $\omega-e$ parameter plane beside ordered domains chaotic regions appeared as well, where in long terms the motion of the transiting Earth-like planet may become unstable. Assuming that chaotic behaviour for an observed transiting planet is unlikely, we could determine an upper limit for the eccentricity, and a lower limit for the inclination. 

In a future work we plan to extend our studies to investigate systematically the stability structure of the $(\omega-e)$ parameter plane for various values of the giant planet's semi-major axis, eccentricity, and inclination. Since the mass of the hosting star and several orbital parameters of the giant planets are known only with a limited accuracy from the radial velocity observations (see for instance Ford et al. 2005), we also plan to follow the propagation of these errors through the method presented in this paper. In our future investigations we intend to consider cases of more massive transiting planets as well. 
 
\acknowledgements  This work has been supported by the Hungarian Scientific Research Fund (OTKA) under the grants D048424 and T043739. The author thanks the valuable help of Dr. C. Beaug\'e and Prof. B. \'Erdi in improving the ma\-nu\-script. The useful comments of the unknown referees are also acknowledged.

\end{article}
\end{document}